\documentclass{kapproc} % Computer Modern font calls

\usepackage{procps} 

\usepackage[dvips]{graphicx}

\upperandlowercase

\kluwerbib % will produce this kind of bibliography entry:

\begin{document}

\articletitle[A new deep SCUBA survey of gravitationally lensing clusters]
{A new deep SCUBA survey of \\ gravitationally lensing clusters}

\author{Kirsten K.\ Knudsen and Paul P.\ van der Werf}

%% affil, email, and abstract are optional
\affil{Leiden Observatory\\
P.O. Box 9513, 2300 RA Leiden\\
The Netherlands}
\email{kraiberg@strw.leidenuniv.nl}

%% optional, to supply a shorter version of the title for the running head:
%%\chaptitlerunninghead{}

\anxx{Knudsen\, Kirsten K.}
\anxx{van der Werf\, Paul P.}

\begin{abstract}
We have conducted a new deep SCUBA
survey, which has targetted 12 lensing galaxy clusters and one 
blank field.  In this survey we have detected several sub-mJy sources after
correcting for the gravitational lensing by the intervening clusters. 
We here present the preliminary results and point out two highlights.
\end{abstract}


\section{Introduction}

The extragalactic background light determined for the whole spectrum
shows that from the structures in the universe most
of the energy is emitted in the IR and in the optical/UV (e.g., 
Fixsen et al., \cite{fixsen}; Madau \& Pozzetti \cite{mp00}).  The latter
is the stellar light, while the former is the processed light, optical/UV
photons absorbed by dust and re-emitted in the IR.  The total energy
output at the wavelengths ranges are comparable.  Studies resolving
the IR background have found that it is
powered by a relatively small number of dusty star forming galaxies.
These galaxies have large star formation rates, $100-3000\,{\rm M_{\odot}
yr^{-1}}$, and are very luminous in the IR, typically
$10^{11-13}\,{\rm L_\odot}$, thus they are 
a.k.a.\ (ultra-)luminous IR galaxies.

As half of the cosmic star formation appears to be hidden behind dust,
studies of the high-$z$ universe in the IR are imperative.  The redshifted
far-IR emission is observable at the submillimetre (submm) wavelengths with
ground based telescopes and instrumentation.  An important aspect of
submm cosmo\-logy is the negative $k$-correction, which allows
observations of objects out to a redshift of 10.  
The observed flux density is approximately constant with redshift, which 
is of great benefit for detection experiments, but as a result it does 
not give an indication of redshift. 
Even if
the source counts are known, different models with largely different
redshift distributions can be used to reproduce the source counts 
(Blain et al.\ \cite{blain}).
As a consequence, to study the star formation history of these objects
the redshift distribution must be determined by measuring the redshift
of the individual objects.   The negative $k$-correction is unique to
only a few wavelengths.  At most other wavelengths the positive
$k$-correction makes it difficult to observe objects at high redshift.
This of course affects the follow-up and identification studies
the submm sources.

\section[A new deep SCUBA survey]
{A new deep SCUBA survey}

We have conducted a survey with SCUBA (Submillimeter Common User Bolometer
Array; Holland et al.\ \cite{holland}), mounted at the J.C.~Maxwell Telescope, 
Hawaii, aimed at studying the faint submm population.
We have targetted 12 galaxy cluster fields and one blank field.
The clusters are strongly gravitationally lensing and have redshifts
between 0.03 and 0.88.  The cluster fields have been observed to a depth 
of $1\sigma_{\rm rms}
\sim 1-2\,{\rm mJy/beam}$ (one field shallower).  55 sources have been
detected in those fields.
The blank field is the NTT Deep Field, which was
observed to $1\sigma_{\rm rms} \sim 1\,{\rm mJy/beam}$.
Here 5 sources were detected.
The total area surveyed is $70\,{\rm arcmin}^{2}$. 
The analysis and source extraction involved Mexican Hat Wavelets
(Cayon et al.\ \cite{mhw}) and Monte Carlo simulations 
(Knudsen et al.\ {\it in prep.}, Barnard et al.\ {\it in prep}). 

\section[Lensing and counts]{Lensing and counts}

\begin{figure}[ht]
\center{\includegraphics[height=6cm]{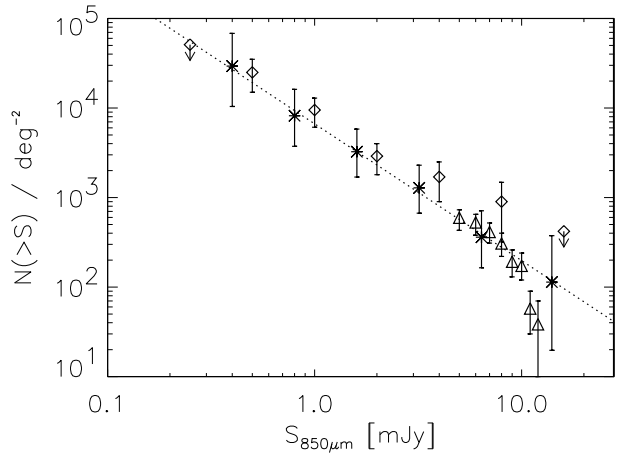}}
\caption{The cumulative $850\,\mu{\rm m}$ source counts
based on half of our survey.  The {\em asterisks} represents this work. 
For comparison the counts from 
Smail et al.\ (\cite{smail02}) ({\em diamonds}) and from Scott et al.\ 
(\cite{scott}) ({\em triangles}) have been included.  
The {\em dotted line} shows the best fit to our data points. 
\label{fig:counts}
}
\end{figure}

The gravitational lensing by the foreground clusters amplifies the 
background sources. Furthermore, it magnifies the regions behind the 
clusters.
As a result the area surveyed in the source
plane is 2-3 times smaller than the area seen in the image plane.  The
effect of lensing is only of benefit when the counts are steep, which they
are in the case of submm sources.  The lensing moves the confusion limit
to fainter flux density levels, hence allowing us to reliably pick out
the faint sources.   Most of the previous surveys made in the submm
have targetted blank fields, and thereby studied the bright and intermediate
population (e.g., Scott et al., \cite{scott}; Webb et al., \cite{webb1}).
Only a limited number of surveys have been able to study
the faint population 
(e.g., Smail et al., \cite{smail98}). 
In our survey, after correcting for the
gravitational lensing, we have been able to observe objects with sub-mJy
fluxes and survey an area large enough for us to study the counts
of the faint population.  
The preliminary number counts based on half of our survey are shown 
in Fig.~\ref{fig:counts}.  Tentatively we find a slope $\alpha = 1.5$, 
when assuming a power law $N(>S) \propto S^{-\alpha}$.  This is 
comparable to other surveys.

\section[ID's]{Follow-up and Identifications}

\begin{figure}[ht]
\narrowcaption{In this mosaic are shown three examples of submm sources 
with one or more very or extremely red objects nearby.  
The contours represent the 
$850\,\mu{\rm m}$ signal-to-noise levels (3,4,5,6 and 7). 
\label{fig:redobjs}
}
\vskip-3.0cm
\includegraphics[width=6.4cm]{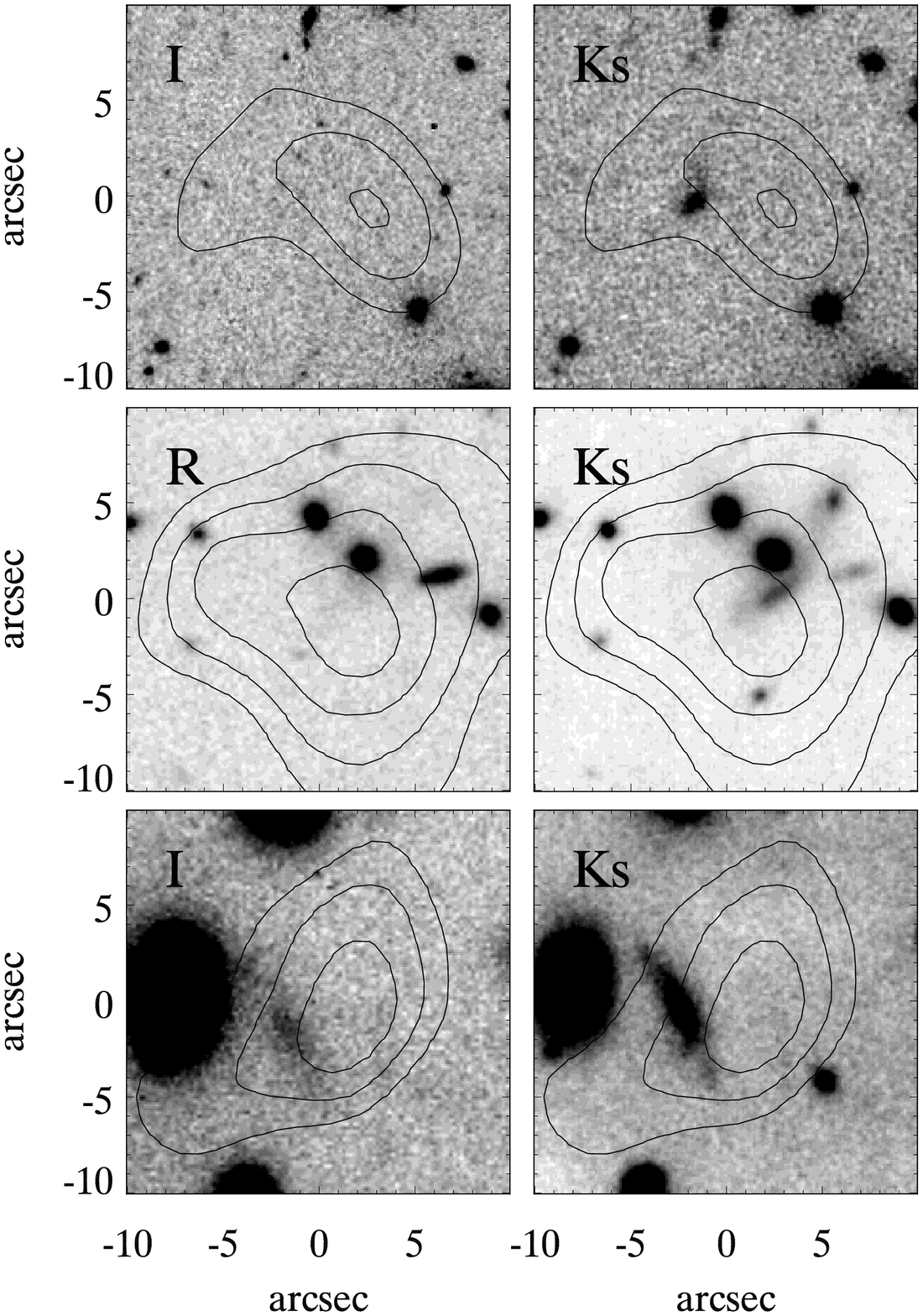}
\end{figure}

We have done follow-up observations with ISAAC at the VLT, obtaining
deep $Ks$ images (limiting magnitude $\sim 21.5\,{\rm mag}$ (Vega)).
This has been combined with archival, deep optical data from HST and VLT, 
and when possible also with observations at other wavelengths (radio, mid-IR).
A substantial fraction of the plausibly identified submm sources 
turn out to be very or extremely red objects (the EROs are often 
defined to have $I-K > 4$). 
Remarkably, in many cases, counterparts identified as red objects 
have close neighbours which are also red objects.
Examples are shown in Fig.~\ref{fig:redobjs}.

We finally point out two highlights of the project: 
a multiple imaged galaxy and a type-1 quasar.

The multiple imaged galaxy is found in the field of A2218 (van der Werf 
et al.\ {\it in prep}).  
Three submm sources were identified with three images of the same very 
red galaxy, SMM\,J16359+6612, detected both in optical and near-IR.  
The submm source is detected both at
$850\,\mu{\rm m}$ and $450\,\mu{\rm m}$ with the same colour for all
three images.  The magnification factor in total for all three images is 40. 
The lensing corrected flux density is $f_{850} = 0.9\,{\rm mJy}$.
Based on detailed lensing models the galaxy has an estimated redshift
of 2.5.   Its spectral energy distribution is similar to that of 
Arp220 and HR10, though about a
factor 5 fainter.

A quasar was identified in the field of A478.  The submillimetre source,  
SMM\,J04135+10277, has
$f_{850} = 25\,\mu{\rm m}$ and $f_{450} = 55\,\mu{\rm m}$ -- these fluxes
should be corrected with a magnification factor of 1.3.  This is the
brightest source in the entire survey and also bright for other similar
surveys.  The quasar has a redshift of 2.837.  
CO has been detected at
$z=2.84$ (Hainline et al, in prep).  An analysis of the SED suggests that
the quasar has a larger submm/far-IR emission than other quasars.
The IR luminosity is $L_{IR} = 3\cdot 10^{13}\,{\rm L}_\odot$.  An
analysis of the optical spectrum suggests that the viewing angle from
the relativistic beam is large, which can have consequences for the
identification of other submm sources (Knudsen et al., \cite{knudsen}).

\end{document}